\documentclass[vecphys]{svmult}
\usepackage{graphicx}
\usepackage{amsmath,amsfonts}

\DeclareGraphicsExtensions {.jpg,.pdf,.ps}

\def\be{\begin{equation}}
\def\ee{\end{equation}}
\def\bea{\begin{eqnarray}}
\def\eea{\end{eqnarray}}
\def\p{\partial}
\def\gcc{g cm$^{-3}$}

\title*{3D meshfree magnetohydrodynamics}
\titlerunning{Meshfree MHD}
\author{Stephan Rosswog\inst{1}\and
Daniel Price\inst{2}}
\institute{Jacobs University Bremen, Campus Ring 1, D-28759 Bremen, Germany 
\texttt{s.rosswog@jacobs-university.de}
\and School of Physics, University of Exeter, Stocker Rd, Exeter EX4 4QL, UK
\texttt{dprice@astro.ex.ac.uk}}

\begin{document}
\maketitle


\begin{abstract}
We describe a new method to include magnetic fields into smooth particle hydrodynamics.
The derivation of the self-gravitating hydrodynamics equations from a variational 
principle is discussed in some detail. The non-dissipative magnetic field evolution
is instantiated by advecting so-called Euler potentials. This approach enforces
the crucial $\nabla\cdot\vec{B}=0$-constraint by construction. These recent developments are
implemented in our three-dimensional, self-gravitating magnetohydrodynamics code
MAGMA. A suite of tests is presented that demonstrates the superiority of this new 
approach in comparison to previous implementations.
\end{abstract}


\section{Specific astrophysical requirements}
\begin{figure}
\centering
\hspace*{-3cm}\includegraphics[height=0.7\textheight]{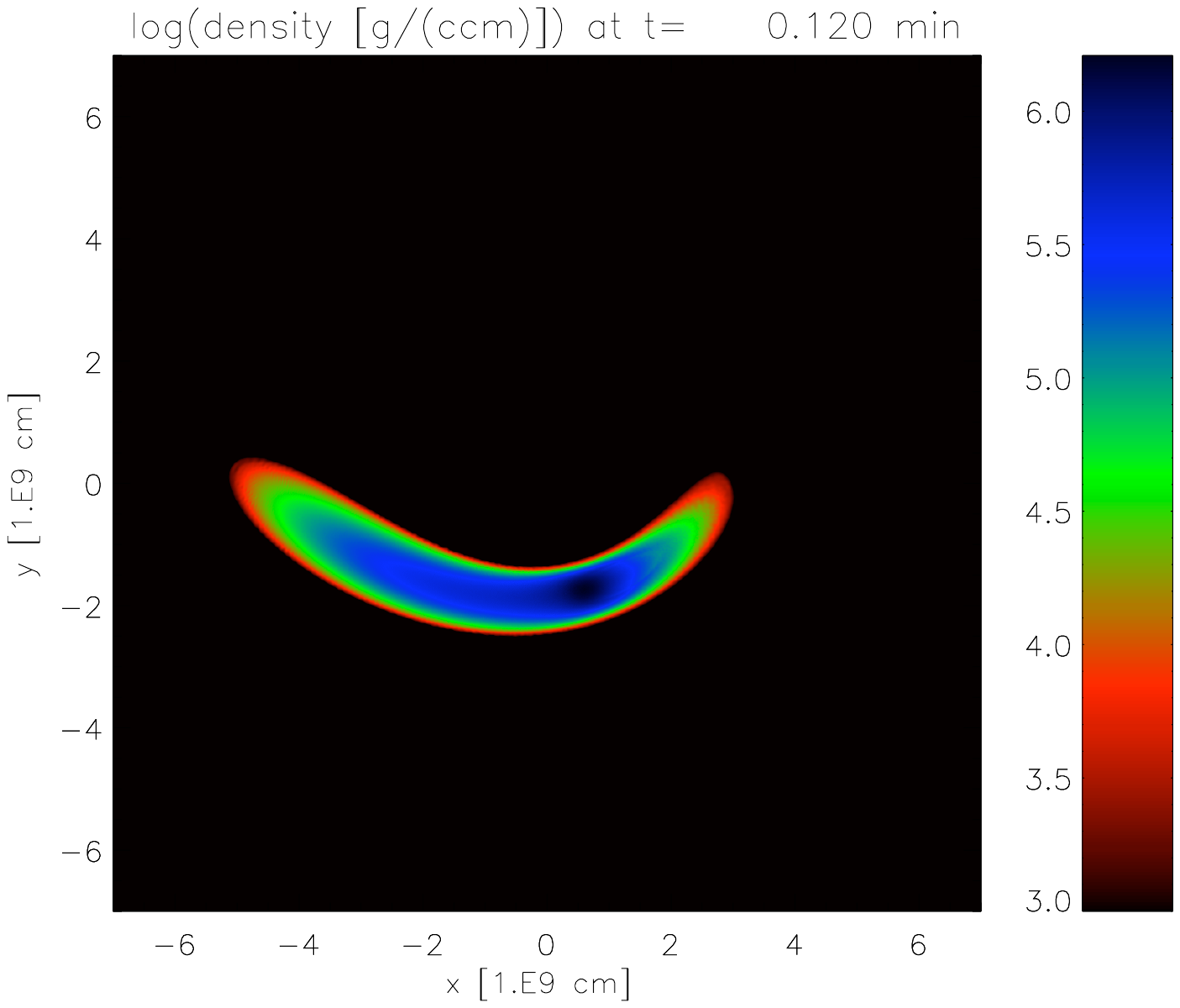}\hspace*{-2.5cm}\includegraphics[height=0.7\textheight]{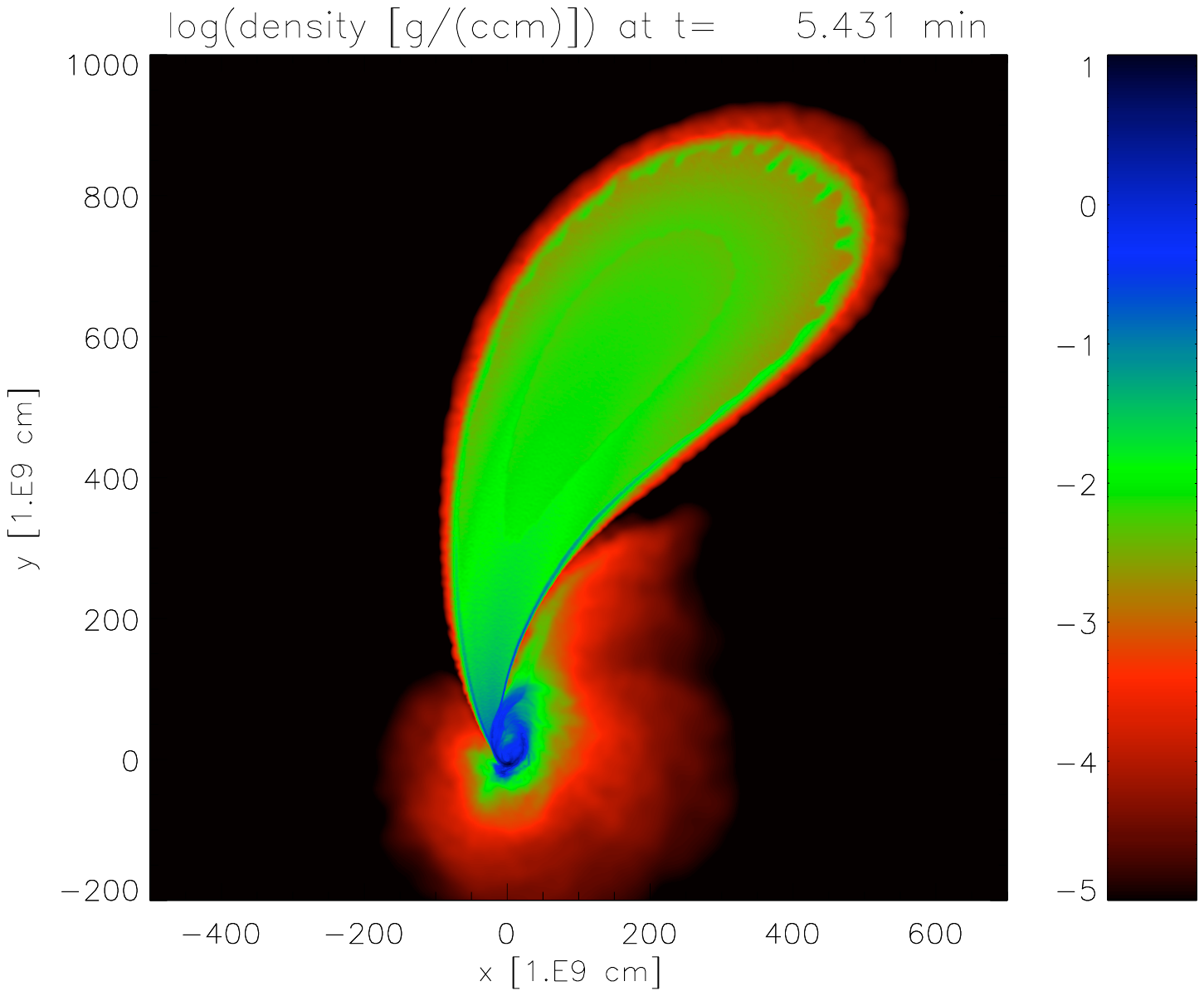}
\caption{Two snapshots from the simulation of the tidal disruption
of a white dwarf by a black hole.}
\label{StephanRosswog::fig:1}       
\end{figure}
Astrophysical simulations have their specific requirements which differ in many respects 
from those of other branches of computationally intense research fields.
The dynamics of self-gravitating gas masses plays a prominent role throughout astrophysics, 
but it is usually only one of several ingredients and it is often necessary to account 
for additional physical processes such as radiative transfer, nuclear burning or the evolution 
of magnetic fields to address questions of astrophysical interest. These additional 
processes often involve intrinsic length and time scales 
that are dramatically different from those of the gas dynamical processes making astrophysical 
problems prime examples of multi-scale and multi-physics challenges.\\
Since fixed boundaries are usually absent, flow geometries are determined by the interplay 
between different physical processes such as gas dynamics and (self-)gravity which often leads to
complicated, dynamically changing flow geometries. Therefore, many problems require flexible 
numerical schemes such as adaptive mesh refinement or completely meshfree, Lagrangian methods. 
Each of these methods has its stengths and weaknesses and the choice of the best-suited method 
can usually save a tremendous amount of effort.\\
An astrophysical example of such an intrinsic multi-scale and multi-physics problem is shown in 
Fig.~\ref{StephanRosswog::fig:1}\footnote{Astrophysical implications of this topic are discussed in
\cite{rosswog08a,rosswog08c}, details of the numerics can be found in \cite{rosswog08b}}. 
It shows two snapshots from a simulation of the tidal disruption of
a white dwarf star by a black hole. The initially spherical star becomes strongly distorted while
passing the black hole (left panel), it is heavily compressed compressed and shock-heated which triggers 
very rapid nuclear reactions whose energy release leads to the thermonuclear explosion of the 
white dwarf. In order to follow this process for each of the computational fluid particles a 
nuclear network \cite{Hix98} is evolved on-the-fly together with the hydrodynamics.\\
In many astrophysical problems the numerical conservation of physically conserved quantities 
determines the success and the reliability of a numerical simulation. Consider, for example, a
molecular gas cloud that collapses under the influence of its own gravity to form stars.
If in the simulation angular momentum is artificially dissipated, say due to too coarse a 
mesh discretization, a collapsing, self-gravitating portion of gas may form just a single 
stellar object instead of a multiple system of stars and it will thus produce a qualitatively wrong 
result. The ``exact''\footnote{``Exact'' means up to possible effects from the numerical 
integration of the resulting ODEs or from using approximative forces, say from a tree or 
some other Poisson-solver.} conservation of mass, energy, linear and angular momentum is --besides
its natural adaptivity-- one of the main strengths of the smoothed particle hydrodynamics (SPH) method.
This exact conservation can be  ``hardwired'' into SPH's evolution equations via symmetries in the 
fluid particle indices. Originally this was done --successfully, but somewhat arbitrarily--
by hand\cite{hernquist89,benz90a,monaghan92}, but more recently it was shown 
\cite{speith98,monaghan01,springel02,monaghan02} how the correct symmetries follow
elegantly and stringently from a discretized fluid Lagrangian and the Euler-Lagrange variational principle.\\
In the following, we will review the derivation of the self-gravitating SPH equations from a Lagrangian, see
Sect.\ref{StephanRosswog:sec:ideal_SPH}. We will also discuss in detail how to implement magnetic fields 
via so-called Euler potentials, see Sect.~\ref{StephanRosswog:sec:ideal_MHD}. This 
approach is similar to evolving a vector potential and  enforces the crucial 
$\nabla\cdot\vec{B}=0$-constraint which otherwise poses a severe challenge for particle methods.
These new develoments are implemented in our self-gravitating, three-dimensional magnetohydrodynamics code
MAGMA, which is described in Sect.~\ref{StephanRosswog:sec:MAGMA}.

\section{Guiding principles}
\subsection{Ideal smoothed particle hydrodynamics (SPH)}
\label{StephanRosswog:sec:ideal_SPH}
The smoothed particle hydrodynamics method (SPH) had originally been developed in the
astrophysical context to simulate the formation of stellar binary systems
via fission\cite{lucy77} and the structure of non-sperical stars\cite{gingold77}.
While the initial 3D simulations used 80 (Gingold and Monaghan) and 100 SPH particles (Lucy)
today's state of the art cosmological SPH simulations have reached particle numbers 
in excess $10^9$, see e.g. \cite{gottloeber06}. This is only in part due to the 
increase in hardware performance, also the simulation techniques (in particular the treatment 
of self-gravity) have become continuously more sophisticated and much effort has been invested
to parallelize 3D codes on various computing platforms.
Also the formulation of the SPH equations has come a very long way from the initial 
straight-forward discretisation of the Lagrangian gas dynamics equations to its most recent
formulation that follows stringently from a discretized ideal fluid Lagrangian.\\
Here we will give a brief overview over an older SPH-formulation, but we will mainly focus on 
an approach that is based on a derivation from a discretised Lagrangian. This latter approach
naturally introduces so-called ``grad-h'' terms that result from changes in the smoothing lengths 
of the SPH particles.

\subsubsection{``Vanilla Ice'' SPH}
The approximation of function values and derivatives via a kernel summation is at the heart
of SPH. If the values of a function $f$ are known at a set of discrete points (``particles'') 
labelled by $b$, the SPH approximation of the function $f$ at position $\vec{r}$ is given
by \cite{benz90a,monaghan92,monaghan05}\footnote{Note that we do not specify at this point
which $h$ is used. For this ``vanilla ice'' SPH the $h$ that enters the kernel should be a 
symmetric combination of the smoothing lengths of the involved particles. This will be explained 
in more detail below. For simplicity, we are omitting the subscript $h$ in
  what follows. We also drop the distinction between the function to be
  interpolated and the interpolant, i.e. we use the same symbol $f$ on both
  sides of the following equation.}
\be
f(\vec{r})= \sum_b \frac{m_b}{\rho_b} f_b W(\vec{r} - \vec{r}_b,h),
\label{StephanRosswog:eq:basic:sum_interpol}
\ee  
where $m_b$ is the (usually constant) particle mass, $\rho_b$ is the mass density and $W$ is 
a kernel function whose width is determined by the smoothing length $h$. Essentially
all astrophysical SPH codes use the cubic spline kernel suggested in \cite{monaghan85}. 
Kernel functions with compact support are preferable since they restrict the SPH-summations to
a local set of neighbours. For the conservation properties it is convenient to have ``radial'' kernels,
\be
W(\vec{r} - \vec{r}_b,h)= W(||\vec{r} - \vec{r}_b||,h),\label{StephanRosswog:eq:basic:radial_kernel}
\ee
so that 
\be
\nabla_a W_{bk}= \nabla_b W_{kb} (\delta_{ba}-\delta_{ka}),\label{StephanRosswog:eq:basic:nabla_a_W_bk}
\ee
and
\be
\nabla_a W_{ab}= \frac{\p W_{ab}}{\p r_{ab}} \; \hat{e}_{ab},\label{StephanRosswog:eq:basic:nabla_a_W_ab_ehat}
\ee
where $\vec{r}_{ab}= \vec{r}_a - \vec{r}_b$, $r_{ab}= ||\vec{r}_{ab}||$, $W_{ab}= W(\vec{r}_{ab},h)$ and $\hat{e}_{ab}=
\vec{r}_{ab}/r_{ab}$. This immediately leads to
\be
\nabla_a W_{ab}= - \nabla_b W_{ab}\label{StephanRosswog:eq:basic:nabla_a_W_ab_asym}
\ee
and 
\be
\frac{dW_{ab}}{dt}= \vec{v}_{ab} \cdot \nabla_a W_{ab},\label{StephanRosswog:eq:basic:dWdt}
\ee
with $\vec{v}_{ab}= \vec{v}_a - \vec{v}_b$ being the velocity difference between particle $a$ and $b$.\\
Eq.~(\ref{StephanRosswog:eq:basic:sum_interpol}) can be applied in particular to the mass density itself which 
then reads
\be
\rho(\vec{r})= \sum_b m_b W(\vec{r} - \vec{r}_b,h).
\label{StephanRosswog:eq:basic:sum_rho}
\ee
The gradient of a function is approximated in SPH by taking the exact derivative of the approximant:
\be
\nabla f(\vec{r})= \sum_b \frac{m_b}{\rho_b} f_b \nabla W(\vec{r} -
\vec{r}_b,h). 
\label{StephanRosswog:eq:basic:nabla_sum}
\ee
The most straightforward and historically first taken approch is to apply this set of rules
to the Lagrangian form of the ideal hydrodynamics equations:
\bea
\frac{d \rho}{dt}&=& - \rho \nabla \cdot \vec{v}, \label{StephanRosswog:eq:basic:drho_dt}\\
\frac{d\vec{v}}{dt}&=& - \frac{\nabla P}{\rho} + \vec{f},\label{StephanRosswog:eq:basic:Euler_eq}\\
\frac{du}{dt}&=& \frac{P}{\rho^2} \frac{d\rho}{dt}= -\frac{P}{\rho} \nabla \cdot
\vec{v}, \label{StephanRosswog:eq:basic:Lag_energy_equation}
\eea
which express the conservation of mass, momentum and energy. Here,  
$P$ is the thermodynamic pressure, $\vec{f}$ abbreviates body forces and $u$ is the thermal 
energy per mass.\\
To briefly illustrate the dependence of conservation on the symmetry of the particle indices let
us apply Eq.~(\ref{StephanRosswog:eq:basic:nabla_sum}) straightforward to the pressure gradient in 
Eq.~(\ref{StephanRosswog:eq:basic:Euler_eq}) (and assume vanishing body forces) to obtain
\be
\frac{d\vec{v}_a}{dt}= -\frac{1}{\rho_a} \sum_b \frac{m_b}{\rho_b} P_b
\nabla_a W_{ab}
\ee
for the acceleration of particle $a$.
This form solves the Euler equation to the order of the method, but it does
not conserve the total momentum. Consider the force that particle $b$ exerts on particle $a$
\be
\vec{F}_{ba}= \left( m_a \frac{d\vec{v}_a}{dt} \right)_b = -
\frac{m_a}{\rho_a}\frac{m_b}{\rho_b} P_b \nabla_a W_{ab}
\label{StephanRosswog:eq:basic:F_ab}
\ee
and similarly, the force from particle $a$ on $b$ 
\be
\vec{F}_{ab}= \frac{m_a}{\rho_a}\frac{m_b}{\rho_b} P_a \nabla_a W_{ab},
\ee
where we have used Eq.~(\ref{StephanRosswog:eq:basic:nabla_a_W_ab_asym}).
Since in general $P_a \neq P_b$, the sum over all the momentum derivatives,
 $\sum_b d (m_b\vec{v}_b)/dt$, does not vanish
and therefore the total momentum is not conserved.\\
This deficiency can be easily cured by expressing the pressure gradient term as
\be
\frac{\nabla P}{\rho} =  \frac{P}{\rho^2} \nabla \rho+ \nabla \left(\frac{P}{\rho}\right).
\label{StephanRosswog:eq:basic:nabla_P_rho}
\ee
If the gradient formula, Eq.~(\ref{StephanRosswog:eq:basic:nabla_sum}), is applied to 
Eq.~(\ref{StephanRosswog:eq:basic:nabla_P_rho}), the momentum equation reads
\bea
\frac{d\vec{v}_a}{dt}
&=& - \sum_b m_b \left(\frac{P_a}{\rho_a^2} + \frac{P_b}{\rho_b^2} \right)
\nabla_a W_{ab}. 
\label{StephanRosswog:eq:basic:momentum_equation}
\eea
Because the pressure part of the equations is now manifestly symmetric in $a$
and $b$ and $\nabla_a W_{ab}=- \nabla_b W_{ba}$ the forces are now equal and
opposite (``actio= reactio'') and therefore the total momentum is conserved 
by construction, i.e. $\sum_a m_a \frac{d\vec{v}}{dt}=0$.\\
Similarly, the total angular momentum is conserved since the sum of all torques
vanishes:
\bea
\frac{d\vec{L}}{dt}&=& \sum_{a,b} \vec{r}_a \times
\vec{F}_{ba}= \frac{1}{2}\left(\sum_{a,b}\vec{r}_a \times \vec{F}_{ba} +
  \sum_{a,b}\vec{r}_a \times \vec{F}_{ba}\right)\nonumber\\
&=& \frac{1}{2}\left(\sum_{a,b}(\vec{r}_a-\vec{r}_b) \times \vec{F}_{ba}\right)= 0.
\eea
Here the summation indices were relabeled and $\vec{F}_{ab} = - \vec{F}_{ba}$ was used. 
The expression finally vanishes, because the forces between particles act along the line
joining them, see Eq.~(\ref{StephanRosswog:eq:basic:nabla_a_W_ab_ehat}).\\
A suitable energy equation can be constructed from 
Eq.~(\ref{StephanRosswog:eq:basic:Lag_energy_equation}) in a straight forward way.
Start from the (adiabatic) first law of thermodynamics
\be
\frac{du_a}{dt}= \frac{P_a}{\rho_a^2} \frac{d\rho_a}{dt}
\label{StephanRosswog:eq:basic:du_dt_a}
\ee
and insert
\be
\frac{d\rho_a}{dt}= \frac{d}{dt} \left(\sum_b m_b W_{ab}\right)
= \sum_b m_b \vec{v}_{ab} \cdot \nabla_a W_{ab},
\label{StephanRosswog:eq:basic:drho_dt_a}
\ee
where we have used Eq.~(\ref{StephanRosswog:eq:basic:dWdt}), to find
\be
\frac{du_a}{dt}= \frac{P_a}{\rho_a^2} \sum_b m_b \vec{v}_{ab} \cdot \nabla_a W_{ab}.
\label{StephanRosswog:eq:basic:energy_equation_u}
\ee
Together with an equation of state the equations
(\ref{StephanRosswog:eq:basic:sum_rho}), (\ref{StephanRosswog:eq:basic:momentum_equation}) and
(\ref{StephanRosswog:eq:basic:energy_equation_u}) form a complete set of SPH equations.\\
In the previous derivation it was implicitely assumed that derivatives 
of the smoothing lengths can be ignored. In a simulation with strongly changing
geometry, however, it is advisable to locally adapt the smoothing length.
This introduces, in principle, additional terms in the SPH equations. 
The importance of these extra terms depends very much on the exact application
\cite{springel02,rosswog07c}.

\subsubsection{The SPH-equations from a Lagrangian, ``grad-h'' terms}

{\em The Lagrangian and the Euler-Lagrange equations}\\
The SPH equations can be derived by using nothing more than a
suitable Lagrangian, the first law of thermodynamics and a prescription on
how to obtain the density via summation.
The Lagrangian of a perfect fluid \cite{eckart60}
\be
L= \int \rho \left(\frac{v^2}{2} - u(\rho,s) \right) dV,
\ee
with $s$ being the specific entropy, can be SPH-discretized in a straightforward way:
\be
L_{\rm SPH,h}= \sum_b m_b \left(\frac{v_b^2}{2} - u(\rho_b,s_b) \right).
\ee
The discretized equations for the fluid are then found by applying
the Euler-Lagrange equations
\be
\frac{d}{dt} \left(\frac{\partial L}{\partial \vec{v}_a} \right) -
\frac{\partial L}{\partial \vec{r}_a} = 0.\label{StephanRosswog:eq:EL_DGL}
\ee
The term in brackets yields the canonical particle momentum 
\be
\frac{\partial L}{\partial \vec{v}_a}=  m_a \vec{v}_a, \label{StephanRosswog:eq:dL_dv}
\ee
the potential-type second term in the Lagrangian becomes
\bea
\frac{\partial L}{\partial \vec{r}_a}&=& - \sum_b m_b \frac{\partial u(\rho_b,s_b)}{\partial \vec{r}_a}
= - \sum_b m_b \left.\frac{\partial u_b}{\partial \rho_b}\right\vert_s 
\cdot \frac{\partial \rho_b}{\partial \vec{r}_a}. 
\label{StephanRosswog:eq:dL_dx}
\eea
The first derivative can be expressed using the first law of thermodynamics,
$du= P/\rho^2 d\rho$, and therefore
\be
m_a \frac{d\vec{v}_a}{dt}= - \sum_b m_b \frac{P_b}{\rho_b^2} \frac{\partial \rho_b}{\partial \vec{r}_a}.\label{StephanRosswog:eq:mi_dot_vi}
\ee\\

{\em The density, its derivatives and the ``grad-h''-terms}\\
We will now address the aditional terms resulting from variable smoothing
lengths. For a density estimate as 
``local'' as possible we use the smoothing length $h_a$ in 
\be
\rho_a= \sum_b m_b W(r_{ab},h_a).\label{StephanRosswog:eq:Advanced_rho}
\ee
Adaptivity can be reached by evolving the smoothing length according to
\be
h_a= \eta \left(\frac{m_a}{\rho_a}\right)^{1/3},\label{StephanRosswog:eq:h}
\ee
where $\eta$ is a parameter typically in a range between 1.2 and 1.5\cite{price04c}.
Since $\rho_a$ and $h_a$ mutually depend on each other, see Eqs.~(\ref{StephanRosswog:eq:Advanced_rho}) 
and (\ref{StephanRosswog:eq:h}), an iteration is required for consistency.\\ 
If we take the changes of $h$ into account, the Lagrangian time derivative of
the density is given by 
\bea
\frac{d\rho_a}{dt}
                  &=& \sum_b m_b \left\{ \frac{\partial W_{ab}(h_a)}{\partial r_{ab}}
                       \frac{d r_{ab}}{dt} 
                      + \frac{\partial W_{ab}(h_a)}{\partial h_{a}} 
                         \frac{d h_{a}}{dt}\right\}\nonumber \\
                  &=& \sum_b m_b \vec{v}_{ab} \cdot \nabla_a W_{ab}(h_a)
                      + \frac{\partial h_a}{\partial \rho_a} 
                      \frac{d \rho_a}{dt} \sum_b m_b 
                      \frac{\partial}{\partial h_a}W_{ab}(h_a)\nonumber,
\eea
where we have used $d r_{ab}/dt= \hat{e}_{ab} \cdot \vec{v}_{ab}$ and 
Eq.~(\ref{StephanRosswog:eq:basic:nabla_a_W_ab_asym}).
If the $d\rho_a/dt$-terms are collected into the quantity 
\be
\Omega_a\equiv \left(1 - \frac{\partial h_a}{\partial \rho_a} \cdot \sum_b m_b
\frac{\partial}{\partial h_a} W_{ab}(h_a) \right),\label{StephanRosswog:eq:omega_a}
\ee
the time derivative of the density reads
\be
\frac{d\rho_a}{dt}= \frac{1}{\Omega_a} \sum_b m_b \vec{v}_{ab} \cdot \nabla_a
W_{ab}(h_a) \label{StephanRosswog:eq:advanced:drho_dt_omega}.
\ee
This is the generalization of the standard SPH
expression, Eq.~(\ref{StephanRosswog:eq:basic:drho_dt_a}). \\
In a similar way the spatial derivatives can be calculated
\bea
\frac{\p\rho_b}{\p\vec{r}_a}
&=& \sum_k m_k \left\{ \nabla_a W_{bk}(h_b)
   + \frac{\partial W_{bk}(h_b)}{\partial h_b} \frac{\p h_b}{\p\vec{r}_a} \right\}\nonumber \\
&=& \sum_k m_k \nabla_a W_{bk}(h_b)
   + \frac{\partial
     h_b}{\partial \rho_b} \frac{\p \rho_b}{\p\vec{r}_a}
     \sum_k m_k \frac{\partial W_{bk}(h_b)}{\partial h_b} \nonumber,  \eea
or, 
\be
\frac{\p\rho_b}{\p\vec{r}_a}
= \frac{1}{\Omega_b} \sum_k m_k \nabla_a W_{bk}(h_b).\label{StephanRosswog:eq:drho_dx}
\ee\\

{\em The SPH equations with ``grad-h''-terms}\\
\label{StephanRosswog:advanved:sec:grad_h}
Inserting  Eq.~(\ref{StephanRosswog:eq:advanced:drho_dt_omega}) into
Eq.~(\ref{StephanRosswog:eq:basic:du_dt_a}) yields the ``grad-h'' energy equation
\be
\frac{d u_{a,\rm h}}{dt}= \frac{1}{\Omega_a}\frac{P_a}{\rho_a^2}
\sum_b m_b \vec{v}_{ab} \cdot  \nabla_a W_{ab}(h_a) \label{StephanRosswog:eq:grad_h_energy}.
\ee
With the derivative Eq.~(\ref{StephanRosswog:eq:drho_dx}) one can write 
Eq.~(\ref{StephanRosswog:eq:mi_dot_vi}) as
\bea
m_a \frac{d\vec{v}_a}{dt}= - \sum_b m_b \frac{P_b}{\rho_b^2} \nabla_a \rho_b
= - \sum_b m_b \frac{P_b}{\rho_b^2} \left(\frac{1}{\Omega_b} \sum_k m_k \nabla_a W_{bk}(h_b)\right).
\eea
With Eq.~(\ref{StephanRosswog:eq:basic:nabla_a_W_bk}), the above equation becomes
\bea
m_a \frac{d\vec{v}_a}{dt} &=&- \sum_b m_b \frac{P_b}{\rho_b^2} \frac{1}{\Omega_b} 
\sum_k m_k \nabla_b W_{kb}(h_b) \; (\delta_{ba}-\delta_{ka})\nonumber\\
&=&- m_a \sum_b m_b 
\left(
\frac{P_a}{\Omega_a \rho_a^2} \nabla_a W_{ab}(h_a) + 
\frac{P_b}{\Omega_b \rho_b^2} \nabla_a W_{ab}(h_b)
\right),
\eea
i.e. the final momentum equation reads
\be
\frac{d\vec{v}_{a,\rm h}}{dt}= - \sum_b m_b 
\left(
\frac{P_a}{\Omega_a \rho_a^2} \nabla_a W_{ab}(h_a) + 
\frac{P_b}{\Omega_b \rho_b^2} \nabla_a W_{ab}(h_b)
\right).\label{StephanRosswog:eq:grad_h_momentum}
\ee
Together with the density equation, Eq.~(\ref{StephanRosswog:eq:basic:sum_rho}),
and an equation of state, Eqs.~(\ref{StephanRosswog:eq:grad_h_energy}) and
(\ref{StephanRosswog:eq:grad_h_momentum}) form a complete set of ``grad-h'' SPH-equations.\\

{\em Self-gravity and gravitational softening}\\
\label{StephanRosswog:sec:grav}
The variational concept can also be applied to derive the gravitational forces 
including softening in a self-consistent way\cite{price07a}.
If gravity is taken into account, a gravitational part has to be added 
to the Lagrangian, $L_{\rm SPH}= L_{\rm SPH,h} + L_{\rm SPH,g}$ with
\be
L_{\rm SPH,g}= - \sum_b m_b \Phi_b, \label{StephanRosswog:eq:Lgrav}
\ee
where $\Phi_b$ is the potential at the particle position $b$,
$\Phi(\vec{r}_b)$. The potential $\Phi$ can be written as a sum 
over particle contributions
\be
\Phi (\vec{r})= - G \sum_b m_b \phi(|\vec{r}-\vec{r}_b|,h), 
\label{StephanRosswog:eq:phi_sum_over_particles}
\ee
and it is related to the matter density by Poisson's equation
\be
\nabla^2 \Phi= 4 \pi G \rho. \label{StephanRosswog:eq:poisson}
\ee
If we insert the sum representations of both the potential, 
Eq.~(\ref{StephanRosswog:eq:phi_sum_over_particles}), and the density, Eq.~(\ref{StephanRosswog:eq:basic:sum_rho}),
into the Poisson equation, Eq.~(\ref{StephanRosswog:eq:poisson}), we obtain a relationship
between the gravitational softening kernel, $\phi$, and the SPH-smoothing
kernel $W$:

\be
W(|\vec{r}-\vec{r}_b|,h)= -\frac{1}{4 \pi} \frac{\partial}{\partial r} 
\left( r^2 \frac{\partial}{\partial r} \phi(|\vec{r}-\vec{r}_b|,h)\right).
\ee
Here we have used that both $\phi$ and $W$ depend only radially on the
position coordinate.\\
Applying the Euler-Lagrange equations, Eq.~(\ref{StephanRosswog:eq:EL_DGL}), 
to $L_{\rm grav,g}$ yields the particle acceleration due to gravity  \cite{price07a}
\bea
\frac{{d\vec{v}}_{a,\rm g}}{dt}&=&  
-G \sum_b m_b \left[ \frac{\phi'_{ab}(h_a)+\phi'_{ab}(h_b)}{2}\right]
\hat{e}_{ab} \nonumber\\
&-& \frac{G}{2} \sum_b m_b 
\left[ 
\frac{\zeta_a}{\Omega_a} \nabla_a W_{ab}(h_a)+
\frac{\zeta_b}{\Omega_b} \nabla_a W_{ab}(h_b)\label{StephanRosswog:eq:dv_dt_grav}
\right],  
\eea
where $\phi'_{ab}= \partial \phi/\partial |\vec{r}_a - \vec{r}_b|$.
The first term in Eq.~(\ref{StephanRosswog:eq:dv_dt_grav}) is the gravitational force 
term usually used in SPH. The second term is due to gradients in the 
smoothing lengths and contains the quantities
\be
\zeta_k\equiv \frac{\partial h_k}{\partial \rho_k} \sum_b m_b \frac{\partial
  \phi_{kb}(h_k)}{\partial h_k}
\ee
and the $\Omega_k$ defined in Eq.~(\ref{StephanRosswog:eq:omega_a}). Formally, it looks very
similar to the pressure gradient terms in Eq.~(\ref{StephanRosswog:eq:grad_h_momentum})
with $G \zeta_k/2$ corresponding to $P_k/\rho_k^2$. As $\zeta_k$ is a negative
definite quantity, these adaptive softening terms act against the gas pressure
and therefore tend to increase the gravitational forces.
The explicit forms of $\phi$, $\phi'$ and $\partial \phi/\partial h$ for 
the cubic spline kernel an be found in Appendix A of \cite{price07a}.

\subsection{Ideal magnetohydrodynamics}
\label{StephanRosswog:sec:ideal_MHD}
Magnetic fields pervade the Universe in substantial strengths on all scales\cite{ruediger04}. They 
are observed in intra-cluster media in galaxy clusters \cite{clarke01} as well as in individual galaxies 
\cite{widrow01}. They are thought to be important for the birth of stars \cite{maclow04}, 
they influence the life of stars e.g. via Sun spots or via controlling
the angular momentum evolution during a stellar lifetime\cite{heger05}. Stellar corpses such as 
neutron stars make themselves known via their magnetic field as pulsars, in a 
particular breed of neutron stars, so-called ``magnetars''\cite{thompson93}, the field 
reaches gigantic field strengths of the order $\sim 10^{15}$ Gauss. On the scale of
planets, magnetic fields controle the magnetospheres that can shield the planet from the 
lethal cosmic rays, a fact that has certainly facilitated the evolution of life on our planet.

\subsubsection{Basic equations of ideal MHD}
Magnetohydrodynamics is a one-fluid model for a highly conducting plasma. It assumes
that electromagnetic fields are highly coupled to the electron-ion component so that
if the fields have a typical frequency $\omega$ and wave number $k$, they fulfill 
$\omega \tau_{\rm h} \sim 1$ and $k \lambda_{\rm h} \sim 1$, where $\tau_{\rm h}$ and 
$\lambda_{\rm h}$ are the typical hydrodynamic time and length scales. If
\be
\frac{1}{\beta_{\rm plas}} \left(\frac{r_{\rm L_{\rm i}}}{\lambda_{\rm h}} \right)^2
\ll \left(\frac{m_{\rm i}}{m_{\rm e}}\right)^{1/2}
\left(\frac{\tau_{\rm i}}{\tau_{\rm h}}\right) \ll 1,
\ee
where $\beta_{\rm plas}$ is the ratio between gas and magnetic pressure, $r_{\rm L_{\rm i}}$ the Larmor radius
of the ions, $m_{\rm i}$ and  $m_{\rm e}$ the ion and electron masses and $\tau_{\rm i}$ is the
typical ion collision time, is fulfilled, the plasma can be described by the equations of
{\rm ideal magnetohydrodynamics}\cite{boyd03}:
\bea
\frac{d \rho}{dt}&=& - \rho \nabla \cdot \vec{v}\\
\frac{dv^i}{dt}&=& \frac{1}{\rho} \frac{\partial S^{ij}}{\partial x^j}\label{StephanRosswog:eq:magn_momentum}\\
\frac{du}{dt}&=& -\frac{P}{\rho} \nabla \cdot \vec{v}\\
\frac{d\vec{B}}{dt}&=& - \vec{B} (\nabla\cdot \vec{v}) + (\vec{B}\cdot \nabla) \vec{v},
\label{StephanRosswog:eq:induction}
\eea
where the
magnetic stress tensor is given by
\be
S^{ij}= -P \delta^{ij} + \frac{1}{\mu_0} \left( B^i B^j - \frac{1}{2} B^2
  \delta^{ij} \right)
\ee
and the $B^k$ are the components of the magnetic field strength.
Note that for ideal magnetohydrodynamics only the momentum equation has to be modified,
the energy and the continuity equation are identical to the case of vanishing magnetic 
field. The form of the momentum equation employed here formally accounts for $\vec{B} 
(\nabla \cdot \vec{B})$ terms which are needed for momentum conservation in shocks but
on the other hand can be the cause of numerical instabilities, see \cite{price04a} 
for a detailed discussion.\\
Due to its relative simplicity in comparison to a more sophisticated plasma treatment 
magnetohydrodynamics and in particular ideal magnetohydrodynamics has been employed throughout
a broad range of applications with sometimes not sufficient consideration about its range 
of applicability. Whether the conditions of applicabilty \cite{jackson98,boyd03} really hold
needs to be checked for each problem individually.

\subsubsection{Euler potentials}
\label{StephanRosswog:sec:Euler}
\begin{figure}
\centering
\includegraphics[height=0.3\textheight]{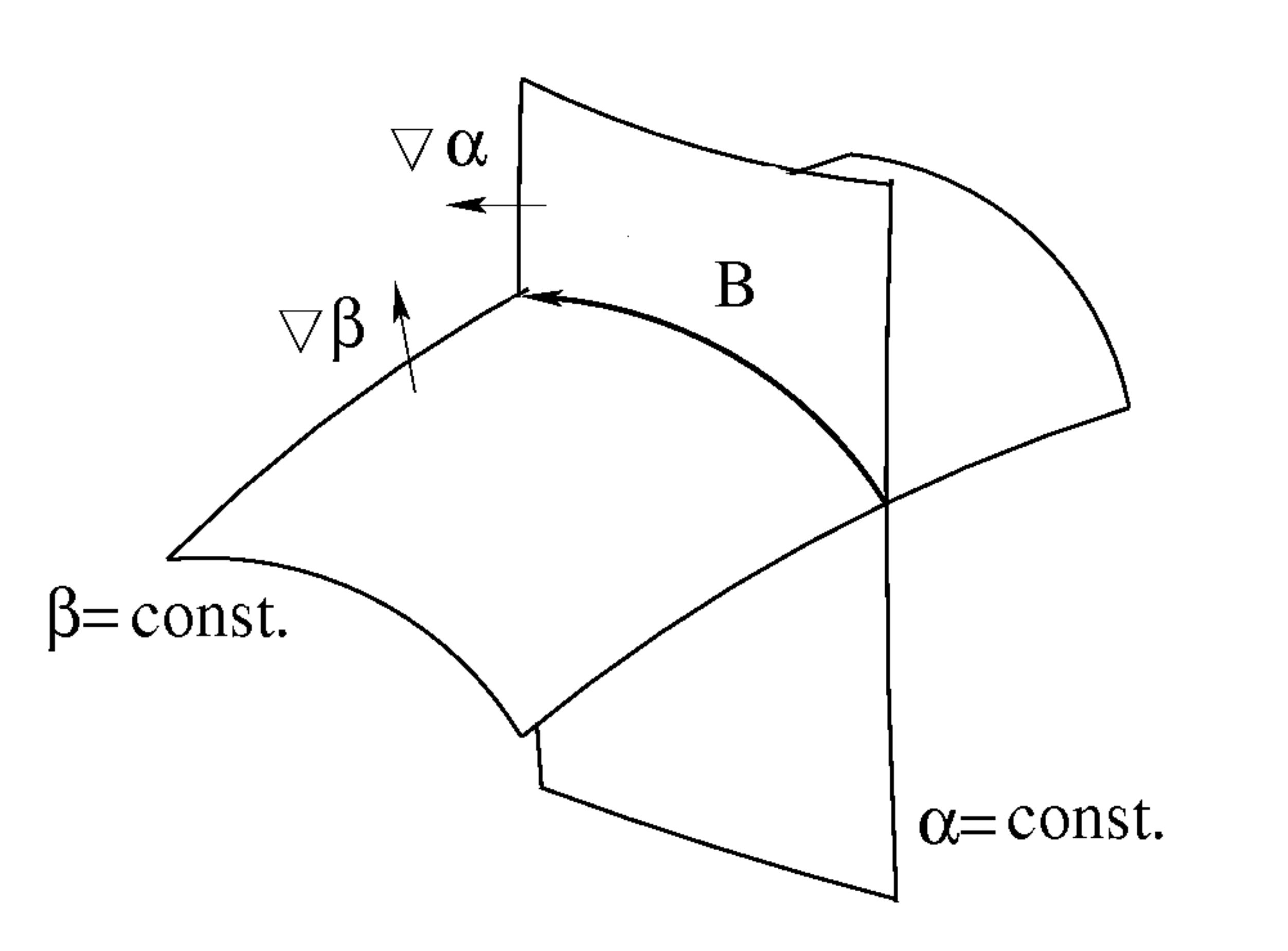}
\caption{The intersection of a plane of constant $\alpha$ with a plane of constant $\beta$
labels a magnetic field line.}
\label{StephanRosswog:fig:Euler_potentials}      
\end{figure}

Being dissipationless the ideal MHD equations are conservative which leads to some
important implications, the most powerful of which is probably the {\em frozen flux
theorem} \cite{alfven51} which states that the magnetic field is carried around by 
the plasma. This kinematic effect is due to the evolution equation of the magnetic field, 
Eq.~(\ref{StephanRosswog:eq:induction}), and represents the conservation of magnetic flux
through a fluid element. In reality, i.e. in the presence of dissipative terms, some slippage
between the magnetic field and the plasma will occur.\\
The idea that the magnetic field lines are carried around by the flow is closely related
to the concepts of Euler potentials \cite{euler1769} which are sometimes also referred to as 
Clebsch variables. For a review on Euler potentials we refer to \cite{stern66,stern70}. 
The basic idea is to present the magnetic field by two scalar variables,
$\alpha$ and $\beta$ such that
\be
\vec{B}= \nabla \alpha \times \nabla\beta. \label{StephanRosswog:eq:B_via_Euler}
\ee
From this definition it is obvious that
\be
\vec{B} \cdot \nabla \alpha = 0 = \vec{B} \cdot \nabla \beta,
\ee
in other words: $\alpha$ and $\beta$ are constant along each field line and can therefore
be used as field line labels. This is graphically represented in 
Fig.~\ref{StephanRosswog:fig:Euler_potentials}.
The frozen flux property of ideal MHD the simply translates into advecting $\alpha$ and $\beta$
with a Lagrangian fluid element:
\be
\frac{d\alpha_a}{dt}= 0 \quad {\rm and } \quad \frac{d\beta_a}{dt}= 0.
\ee
\\
The Euler potentials naturally relate to the magnetic vector potential via
\be
\vec{A} = \alpha\nabla\beta + \nabla \xi
\ee
or
\be
\vec{A} = -\beta\nabla\alpha + \nabla \psi,
\ee
where $\xi$ and $\psi$ are arbitrary smooth functions. It is straightforward to check 
that both of the above forms of the vector potential yield the magnetic field via
\be
\nabla \times \vec{A}= \nabla \alpha \times \nabla\beta = \vec{B}.
\ee
Thus, the $\nabla \cdot \vec{B}=0$-constraint that is otherwise very hard to fulfill in a 
particle method \cite{morris96a,price04c,price05} can be hard-wired into the numerical scheme
by using the advected quantities $\alpha$ and $\beta$ to construct the magnetic field via 
Eq.~(\ref{StephanRosswog:eq:B_via_Euler}). This approach has the additional ease that, as long as 
the magnetic field is not strong enough to substantially influence the dynamics of the plasma,
i.e. in the high-$\beta_{\rm plas}$-case, the evolution of different initial field configurations
can be explored by just re-processing an existing simulation with different initial values for 
$\alpha$ and $\beta$. To find Euler potential pairs for a given magnetic field configuration
is however usually a non-trivial task due to the non-linear nature of 
Eq.~(\ref{StephanRosswog:eq:B_via_Euler}). The Euler potentials for a dipole field are known,
but for more complicated field geometries its usually a challenge to find an analytical expression
for the Euler potentials. There are however numerical procedures to find suitable
pairs of Euler potentials, see e.g. \cite{ho97,peymirat99}.\\
In two dimensions, a magnetic field can be represented by
\be
\alpha= A_{z} \quad \beta= z,
\ee
where $A_z$ is the $z$-component of a vector potential.

\subsubsection{Limitations of the Euler potential approach}
While the Euler potential approach makes some otherwise rather challenging
problems such as magnetic field advection (see below) a trivial task, they
have their own difficulties and limitations.\\
First, the Euler potentials for a given field configuration are not
uniquely determined \cite{stern70}. Assume, for example, that one particular set of
Euler potentials, $\alpha_1$ and $\beta_1$, is known. Then for a second set that 
is a function of the known ones, $\alpha_2= \alpha_2(\alpha_1,\beta_1)$ and 
$\beta_2= \beta_2(\alpha_1,\beta_1)$, one finds
\be
\nabla \alpha_2 \times \nabla \beta_2 = \left(\frac{\p \alpha_2}{\p \alpha_1} 
\frac{\p \beta_2}{\p \beta_1} - \frac{\p \beta_2}{\p \alpha_1} \frac{\p \alpha_2}
{\p \beta_1} \right) \nabla \alpha_1 \times \nabla \beta_1
\ee
and therefore $\alpha_2$ and $\beta_2$ will also be a set of Euler potentials
for the same field  as long as the term in brackets is equal to unity.\\
Second, by their very nature the Euler potentials are restricted to the purely 
non-disspative case and thus they are not immediately suited to treat the case of 
dissipative effects in a plasma.\\
Third, there are restrictions with respect to the magnetic field geometries
that can be represented by Euler potentials. It is, for example, impossible
to represent a linked ploidal and toroidal field. Nevertheless, as will be demonstrated
in Sec.\ref{StephanRosswog:sec:tests}, on a large set of standard MHD-test problems
the Euler potential approach yields excellent results.\\
From a numerical point of view they involve higher-order derivatives,
see Eqs.~(\ref{StephanRosswog:eq:magn_momentum}) and (\ref{StephanRosswog:eq:B_via_Euler})
which is usually numerically challenging. However, as will be shown below, this not 
necessarily has to degrade the accuracy of the solution.

\subsection{Dissipative terms}
\label{StephanRosswog:sec:diss}

In both hydrodynamics and magnetohydrodynamics we are interested in 
principle in the non-dissipative cases, see Sec.~\ref{StephanRosswog:sec:ideal_SPH}
and \ref{StephanRosswog:sec:ideal_MHD}. The corresponding 
equations, however, allow for discontinuous shock solutions which need to be captured 
in order to allow for a physically correct and numerically stable solution.
This can be done by either making use of the analytical solution by 
locally solving a Riemann-type problem or by artificially spreading the 
discontinuities to a numerically resolvable width which means making them 
continuous. This latter artificial viscosity approach is most often used 
in the context of smooth particle hydrodynamics, although Riemann-solver-type 
approaches also do exist \cite{inutsuka02,cha03}.\\
A careful design of artificial dissipation terms is essential to capture
physically correct solutions. This was recently demonstrated at the
example of Kelvin-Helmholtz instabilities \cite{price08a}.
In the design of artificial dissipation terms we are guided by two principles: 
a) we want to use a form of the artificial dissipation equations that is 
oriented at Riemann-solvers \cite{monaghan97} and b) we aim at applying 
dissipative terms only where they are necessary, i.e. near discontinuities, 
and follow in this respect Morris and Monaghan \cite{morris97} who suggested 
to use time dependent dissipation parameters.\\
Based on the analogy with Riemann solvers Monaghan \cite{monaghan97} 
presented a general formulation of dissipative terms. It was noted
that the evolution equations of every conservative quantity should contain
dissipative terms to controle discontinuities. This approach has been applied
to ultra-relativistic \cite{chow97} and magnetohydrodynamic 
shocks \cite{price05}. The ``discontinuity capturing'' term for a variable
$A$ is of the form
\be
\left(\frac{dA}{dt}\right)_{a,\rm diss}= \sum_b m_b 
\frac{\alpha_A v_{{\rm sig},A}}{\rho_{ab}} (A_a-A_b) \hat{e}_{ab} \cdot 
\nabla W_{ab},
\label{StephanRosswog:eq:gen_dissipation}
\ee 
where $\alpha_A$ is a number of order unity that specifies the 
exact amount of dissipation, $v_{{\rm sig},A}$ is an appropriate signal 
velocity and $\rho_{ab}$ the average mass density of particles $a$ and $b$.\\
A comparison with the SPH expression for Laplacians \cite{brookshaw85}
shows that the above equation is really an expression for\cite{price08a}
\be
\left(\frac{dA}{dt}\right)_{a,\rm diss}= \eta \nabla^2 A
\ee 
with $\eta \propto \alpha_A v_{\rm sig} |r_{ab}|$.\\
Following \cite{morris97} the parameter that determines the exact values
of the dissipative parameters, $\alpha_A$, is made time-dependent. This
is put into effect by integrating an additional differential equation
containing both a source term, $S_A$, that indicates the necessity of 
artificial dissipation and a decay term that contains the typical time scale,
$\tau_A$, it takes a particle to pass the discontinuity. The evolution equation
of the dissipation parameter is given by
\be
\frac{d\alpha_{A,a}}{dt}= - \frac{\alpha_{A,a} - \alpha_{\rm min}}{\tau_{A,a}} + S_{A,a},
\ee
where $\alpha_{\rm min}$ is the minimum value to which we allow $\alpha_A$ to decay.
The decay time scale is given by
\be
\tau_{A,a}= \frac{h_a}{C v_{{\rm sig},A}},
\ee
where $C$ is a constant of order unity that is chosen after careful 
numerical experiments at problems with analytically known solutions.

\section{The MAGMA code}
\label{StephanRosswog:sec:MAGMA}
Collisions between stars are very rare events in the solar neighbourhood.
Close to centres of galaxies and globular clusters, however, the number 
densities of stars are higher by up to a factor of $10^6$ \cite{heggie03}
and therefore stellar collisions are very common events. In fact, the 
innermost 0.3 lightyears of our Galaxy can be considered an efficient
``stellar collider''\cite{alexander05}. A different type of encounter
can occur for stellar binary systems that contain compact stellar objects.
If born at close enough separations such systems can be driven towards
merger by the emission of gravitional waves. Although rare per space 
volume these types of encounters release tremendous amounts of 
gravitational energy and are therefore potentially visibly out to 
cosmological distances thereby making huge volumes observationally 
accessible and producing a substantial observational rate.\\
Some of the most exciting astrophysical objects are thought to form 
in such encounters and since both neutron stars and white dwarfs are 
known to be threaded by very large magnetic fields, a careful study of 
such mergers requires the inclusion of magnetic fields and their evolution.

\subsection{Scope and physics modules}
The acronym MAGMA stands for {\em a magnetohydrodynamics code for merger 
applications} and this code has originally been developed for the study of
magnetized neutron stars \cite{price06b,rosswog07b}. A very detailed 
description of this code can be found in \cite{rosswog07c}.\\
For astrophysical studies the code contains several physics modules 
that go beyong the scope of this article and shall only be briefly sketched
here. The interested reader is referred to the astrophysical literature.\\

{\em Equation of state}\\
For the thermodynamic properties of neutron star matter we use
a temperature-dependent relativistic mean-field equation of state 
 \cite{shen98a,shen98b}. It can handle temperatures from 0 to 100 
MeV\footnote{1 MeV corresponds to $1.16 \cdot 10^{10}$ K.}, electron 
fractions from $Y_e$= 0 (pure neutron matter) up to 0.56 and densities 
from about 10 to more than $10^{15}$ \gcc. No attempt is made to 
include matter constituents that are more exotic than neutrons and 
protons at high densities. For more details we refer to 
\cite{rosswog02a}.\\

{\em Neutrino emission}\\
The code contains a detailed multi-flavor neutrino leakage scheme. 
An additional mesh is used to calculate the neutrino opacities that are 
needed for the neutrino emission rates at each particle position.
The neutrino emission rates are
used to account for the local cooling and the compositional changes due to 
weak interactions such as electron captures. A detailed description of the 
neutrino treatment can be found in \cite{rosswog03a}.\\

{\em Self-gravity}\\
The self-gravity of the fluid is treated in a Newtonian fashion.
Both the gravitational forces and the search for the
particle neighbors  are performed with a binary tree that is based 
on the one described in \cite{benz90b}. These tasks are the computationally 
most expensive part of the simulations and in practice they completely
dominate the CPU-time usage.  Forces emerging from the emission of
gravitational waves are treated in a simple approximation. For more details,
we refer to the literature \cite{rosswog00,rosswog02a}.

\subsection{The MAGMA equations}
Here, we will only briefly summarize the implemented equations, the explicit 
forms of all the equations can be found in \cite{rosswog07c}.\\
Instead of explicitely integrating the continuity equation, we calculate the
density via summation as in Eq.~(\ref{StephanRosswog:eq:basic:sum_rho}).
The momentum equation is used in the form
\be
\frac{d\vec{v}_{a,\rm MHD}}{dt}= 
\frac{d\vec{v}_{a,\rm h}}{dt} + 
\frac{d\vec{v}_{a,\rm h, diss}}{dt}+
\frac{d\vec{v}_{a,\rm g}}{dt} + 
\frac{d\vec{v}_{a,\rm mag}}{dt}+
\frac{d\vec{v}_{a,\rm mag, diss}}{dt}
\ee
where $d(\vec{v}_{a,\rm h})/dt$ is given in 
Eq.~(\ref{StephanRosswog:eq:grad_h_momentum}), $d(\vec{v}_{a,\rm g})/dt$
is given in Eq.~(\ref{StephanRosswog:eq:dv_dt_grav}), and the explicit forms of
the dissipative terms,$d(\vec{v}_{a,\rm h, diss})/dt$ and 
$d(\vec{v}_{a,\rm mag, diss})/dt$, can be found in \cite{rosswog07c}. 
The magnetic force term is used in the form
\be
\frac{d\vec{v}_{a, \rm mag}}{dt} \hspace*{-0.1cm} =  -\sum_{b} \frac{m_{b}}
{\mu_{0}} 
\left\{ \frac{B_{a}^{2}/2}{\Omega_{a}\rho_{a}^{2}}\nabla_{a}W_{ab}(h_{a})
       +\frac{B_{b}^{2}/2}{\Omega_{b}\rho_{b}^{2}}\nabla_{a}W_{ab}(h_{b}) 
\right\}\nonumber 
\ee
\begin{equation}
 + \sum_{b} \frac{m_{b}}{\mu_{0}} \left\{ 
\frac{\vec{B}_{a} (\vec{B}_{a}\cdot \overline{\nabla_{a}W_{ab}}) 
- \vec{B}_{b} (\vec{B}_{b}\cdot \overline{\nabla_{a}W_{ab}})}
{\rho_{a} \rho_{b}} \right\},
\label{StephanRosswog:eq:fmorr}
\end{equation}
where the symmetrized kernel gradient 
is given by
\be
\overline{\nabla_a W_{ab}}= \frac{1}{2}
\left[\frac{1}{\Omega_a} \nabla_a  W_{ab} (h_a) + 
\frac{1}{\Omega_b} \nabla_a  W_{ab} (h_b)\right]. 
\ee
The magnetic field is calculated from the Euler potentials\footnote{Note
that the code also allows to evolve magnetic fields according to a more 
straightforward SPH discretisation \cite{price05}.}. Note that another form 
of the magnetic force term is also possible \cite{price05,rosswog07c}. 
The gradients of the Euler potentials are calculated in a way that gradients 
of linear functions are reproduced exactly \cite{price04c,rosswog07c}. To 
handle magnetic shocks artificial dissipation terms were constructed 
according to the ideas outlined in 
Sec.~\ref{StephanRosswog:sec:diss}. They are also applied to the evolution 
of $\alpha_a$ and $\beta_a$. They are not meant to mimic physical dissipation
in any way, their exclusive aim is to keep gradients numerically treatable.\\
The MAGMA energy equation is of the form
\be
\frac{du_{a,\rm MHD}}{dt}= 
\frac{d u_{a,\rm h}}{dt} + 
\frac{d u_{a, AV}}{dt}   +
\frac{d u_{a, C}}{dt},
\ee
where $d(u_{a,\rm h})/dt$ is given in 
Eq.~(\ref{StephanRosswog:eq:grad_h_energy}), the explicit form of the
artificial viscosity term $d(u_{a, AV})/dt$ and the thermal conductivity
term $d(u_{a, C})/dt$ can be found in \cite{rosswog07c}.

\subsection{Tests and benchmarks}
\label{StephanRosswog:sec:tests}
We present here a selection of standard tests used in the hydro- and 
magnetohydrodynamics community to validate numerical schemes. For a more
exhaustive set of benchmarks we refer to \cite{rosswog07c}.

\subsubsection{Hydrodynamics}
{\em 1D: Sod's shock tube}\\
\label{sec:sod}
\begin{figure}
\begin{center}
\includegraphics[height=0.5\textheight,angle=-90]{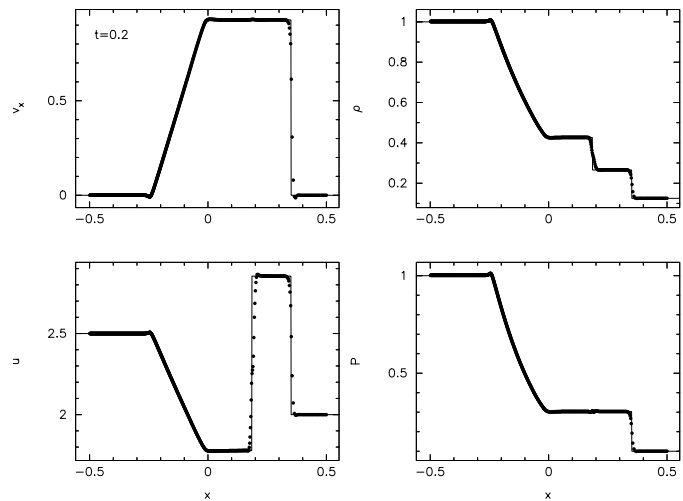}
\caption{Results of the Sod shock tube test in one dimension using 900 
SPH particles setup using unsmoothed initial conditions. Artificial 
viscosity and thermal conductivity are applied to appropriately smooth 
the shock and contact discontinuity respectively. The exact solution is 
given by the solid line. The upper row displays the
velocity (left) and the density (right), the bottom row shows specific
internal energy (left) and the pressure (right).}
\label{fig:sod1}
\end{center}
\end{figure}
As a standard test of the shock capturing capability we show 
the results of Sod's shock tube test \cite{sod78}. 
To the left of the origin, the initial state of the fluid is given by 
[$\rho, P, v_{x}$]$_{\rm L}$ = [1.0,1.0,0.0] whilst to the right of 
the origin the initial state is [$\rho, P, v_{x}$]$_{\rm R}$ = 
[0.125,0.1,0.0] with $\gamma = 1.4$. 
The problem is setup using 900 equal mass particles in one spatial dimension.
Rather than adopting the usual practice of smoothing the initial conditions 
across the discontinuity, we follow \cite{price04c} in using unsmoothed 
initial conditions but applying a small amount of artificial thermal 
conductivity. The results 
are shown in Figure~\ref{fig:sod1}, where the points represent the SPH 
particles. For comparison the exact solution computed using a Riemann solver 
is given by the solid line.\\ 
The shock itself is smoothed by the artificial viscosity term, which in 
this case can be seen to spread the discontinuity over about 6 particles. 
The contact discontinuity is smoothed by the application of artificial 
thermal conductivity which (in particular) eliminates the ``wall heating'' 
effect often visible in numerical solutions to this problem. The exact 
distribution of particle separations in the contact discontinuity seen 
in Figure~\ref{fig:sod1} is related to the initial particle 
placement across the discontinuity.\\
For this test, applying artificial viscosity and thermal conductivity as 
described, we do not find a large difference between the ``grad-$h$''
formulation and other variants of SPH based on averages of the smoothing 
length. If anything, the ``grad-$h$''-terms tend to increase the order of the 
method, which, as in any higher order scheme, tends to enhance oscillations 
which may otherwise be damped, visible in Figure~\ref{fig:sod1} as 
small ``bumps'' at the head of the rarefaction wave (in the absence of 
artificial viscosity these bumps appear as small but regular oscillations 
with a wavelength of a few particle spacings).\\

{\em 3D: Sedov blast wave test}\\
\begin{figure*}
\begin{center}
\includegraphics[height=0.5\textheight,angle=-90]{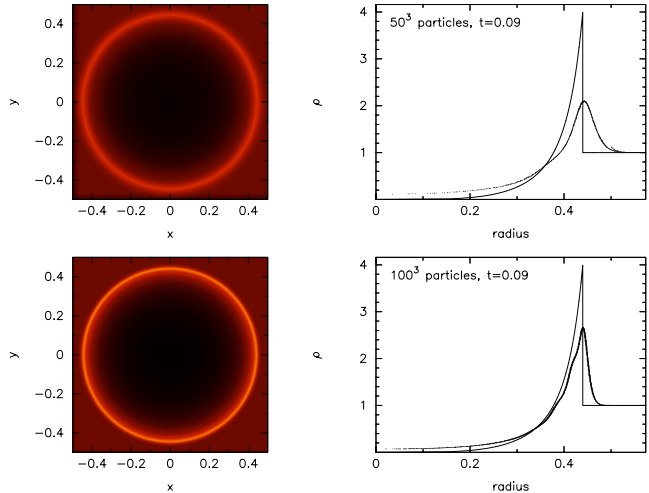}
\caption{Results of the hydrodynamic Sedov blast wave test in 3D at 
$t=0.09$ at resolutions of 125,000 (top) and 1 million (bottom) 
particles respectively. The density and radial position of each 
SPH particle are shown in each case, which may be compared to the 
exact solution given by the solid line.}
\label{fig:sedov}
\end{center}
\end{figure*}
In order to demonstrate that our scheme is capable of handling strong 
shocks in three dimensions, we have also tested the code on a Sedov 
blast wave problem both with, see Sec.~\ref{sec:MHD_blast}, and without
magnetic fields. Without  
magnetic fields the explosion is spherically symmetric, however for 
a strong magnetic field the blast wave is significantly inhibited 
perpendicular to the magnetic field lines, resulting in a compression 
along one spatial dimension. Similar tests for both hydrodynamics and 
MHD have been used by many authors -- for example by \cite{balsara01} 
in order to benchmark an Adaptive Mesh Refinement (AMR) code for MHD 
and by \cite{springel05} in benchmarking the cosmological SPH code GADGET.\\
The hydrodynamic version is set up as follows: The particles are 
placed in a cubic lattice configuration in a three dimensional domain 
$[-0.5,0.5] \times [-0.5,0.5] \times [-0.5,0.5]$ with uniform density $\rho =
1$ and zero  
pressure and temperature apart from a small region $r < R$ near the 
origin, where we initialize the pressure using the total blast wave 
energy $E=1$, ie. $P = (\gamma - 1) E/ (\frac43 \pi R^{3})$. We set 
the initial blast radius to the size of a single particle's smoothing 
sphere $R=2\eta \Delta x$ (where $2$ is the kernel radius, $\eta (= 1.5)$ 
is the smoothing length in units of the average particle spacing as in 
Eq.~(\ref{StephanRosswog:eq:h}) and $\Delta x$ is the initial particle spacing 
on the cubic lattice) such that the explosion is as close to point-like 
as resolution allows. Boundaries are not important for this problem, 
however we use periodic boundary conditions to ensure that the particle 
distribution remains smooth at the edges of the domain.\\
The results shown in Figure~\ref{fig:sedov} at $t=0.09$ have been obtained
with a resolution of 50 and 100 particles$^{3}$ (ie. 125,000 and 1 million 
particles respectively), where we have plotted (left panels) the 
density in a $z=0$ cross section slice and (right panels) the density 
and radial position of each particle (dots) together with the exact 
self-similar Sedov solution (solid line).\\ 
We found that the key to an accurate simulation of this problem in 
SPH is to incorporate an artificial thermal conductivity term due to 
the huge initial discontinuity in thermal energy. The importance of 
such a term for shock problems in SPH has been discussed recently by 
\cite{price04c,price08a}. In the absence of this term the particle distribution 
quickly becomes disordered around the shock front and the radial 
profile appears to be noisy. From Figure~\ref{fig:sedov} we see 
that at a resolution of 1 million particles the highest density in 
the shock at $t=0.09$ is $\rho_{\rm max}=2.67$ whereas for the lower 
resolution run $\rho_{\rm max} = 2.1$, consistent with a factor of 2 
change in smoothing length. Using this we can estimate that a 
resolution of $\sim345^{3} = 41$~million particles is required to 
fully resolve the density jump in this problem in three dimensions. 
Note that the minimum density obtained in the post-shock rarefaction 
also decreases with resolution. Some small-amplitude post-shock 
oscillations are visible in the solution which we attribute to 
interaction of the spherical blast wave with particles in the 
surrounding medium initially placed on a regular (Cartesian) cubic 
lattice.

\subsubsection{Magnetohydrodynamics}

\begin{figure*}
\begin{center}
\includegraphics[height=0.5\textheight,angle=-90]{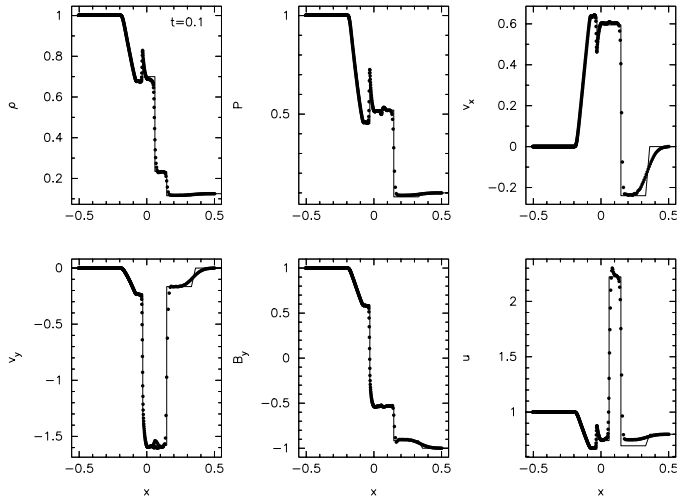}
\caption{Results of the Brio \& Wu MHD shock tube test at $t=0.1$ using 631 particles and the Euler potential formulation. For comparison the numerical solution taken from \cite{balsara98} is given by the solid line. The solution illustrates the complex shock structures which can be formed due to the different wave types in MHD, including in this case a compound wave consisting of a slow shock attached to a rarefaction wave. }
\label{fig:Brio_Wu}
\end{center}
\end{figure*}

{\em 1D: Brio-Wu shock tube test}\\
The magnetic shock tube test of \cite{brio88} has become a standard test
case for numerical MHD schemes that has been widely used by many authors 
to benchmark (mainly grid-based) MHD codes 
\cite{stone92a,dai94,ryu95,balsara98}.
The Brio-Wu shock test is the MHD analogon to 
Sod's shock tube problem that was described earlier, 
but here no analytical solution is known. The MHD Riemann problem 
allows for much more complex solutions than the hydrodynamic case
which can occur because of the three different types of waves (i.e. 
slow, fast and Alfv\'en, compared to just the sound waves in hydrodynamics).
In the Brio-Wu shock test the solution contains
the following components (from left to right in Fig.~\ref{fig:Brio_Wu}): 
a fast rarefaction fan and a slow compound wave consisting of a slow 
rarefaction attached to a slow shock (moving to the left) and a contact 
discontinuity, a slow shock and a fast rarefaction fan (moving to the 
right). It has been pointed out, however, that the stability of the unusual
compound wave may be an artifact of the restriction of the symmetry to one 
spatial dimension whilst allowing the magnetic field to vary in two 
dimensions, \cite{barmin96}.\\
The shown results are obtained using Euler potential formulation. 
Results of this problem using 
Smoothed Particle Magnetohydrodynamics (SPMHD) have been presented 
elsewhere \cite{price04a,price04c}. The Euler potentials show a distinct 
improvement over the standard SPMHD results.
The initial conditions on the left side of the discontinuity are 
$[\rho,P,v_x,v_y,B_y]_{\rm L}= [1,1,0,0,1]$ and
$[\rho,P,v_x,v_y,B_y]_{\rm R}= [0.125,0.1,0,0,-1]$ on the right side. 
The $x-$component of the magnetic field is $B_{x}= 0.75$ everywhere and
a polytropic exponent of $\gamma = 2.0$ is used. Using the Euler potentials
the components are given  
by $\alpha = -B_{y} x$ (equivalent to the vector potential $A_{z}$) and 
$\beta = z$ (or more specifically $\nabla\beta = \hat{\bf z}$) and the 
$B_{x}$ component is treated as an external field which requires adding 
a source term to the evolution equation for $\alpha$. Particles are 
restricted to move in one spatial 
dimension only, whilst the magnetic field is allowed to vary in two 
dimensions (that is, we compute a $v_{y}$ but do not use it to move the 
particles).\\
We setup the problem using 631 equal mass particles in the domain 
$x \in [-0.5,0.5]$ using, as in the hydrodynamic case, purely discontinuous 
initial conditions. Artificial viscosity, thermal conductivity and 
resistivity are applied. The results are shown at $t=0.1$ in 
Figure~\ref{fig:Brio_Wu}. For comparison the numerical solution from 
\cite{balsara98} is given by the solid line (no exact solution exists 
for this problem). The solution is generally well captured by our 
numerical scheme. Two small defects are worth noting. The first is that a 
small offset is visible in the thermal energy -- this is a result of the 
small non-conservation introduced by use of the Morris 
formulation \cite{morris96a} of the magnetic force, 
Eq.~(\ref{StephanRosswog:eq:fmorr}). Secondly, the rightmost discontinuity  
is somewhat over-smoothed by the artificial resistivity term. We attribute 
this to the fact that the dissipative terms involve simply the maximum 
signal velocity $v_{sig}$ (that is the maximum of all the wave types). 
Ideally each discontinuity should be smoothed taking account of it's 
individual characteristic and corresponding $v_{sig}$ (as would occur in 
a Godunov-MHD scheme). Increasing the total number of particles also 
decreases the smoothing applied to this wave.

\begin{figure}
\begin{center}
\includegraphics[height=0.5\textheight,angle=-90]{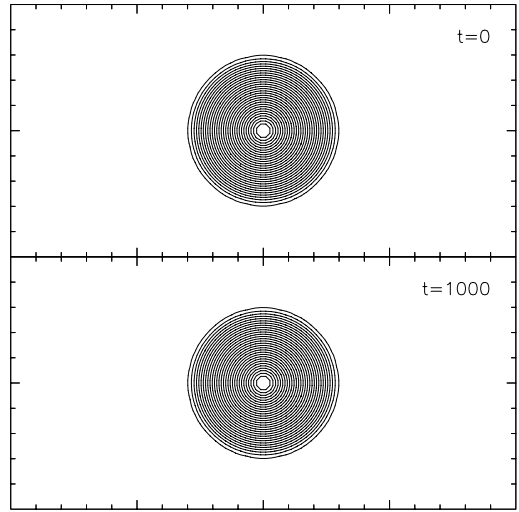}
\caption{Magnetic field lines in the current loop advection test, plotted at
  $t=0$ (top) and after 1000 crossings of the computational domain (bottom).}
\label{fig:jadvect2D}
\end{center}
\end{figure}

{\em 2D: Current loop advection problem}\\
A simple test problem for MHD is to compute the advection of a weak 
magnetic field loop. This test, introduced by \cite{gardiner05} in the 
development of the {\it Athena} MHD 
code\footnote{http://www.astro.princeton.edu/$\sim$jstone/athena.html}, 
presents a challenging problem for grid-based MHD schemes requiring 
careful formulation of the advection terms in the MHD equations. For 
our Lagrangian scheme, this test is straightforward to solve 
which strongly highlights the advantage of using a particle method 
for MHD in problems where there is significant motion with respect 
to a fixed reference frame.\\
We setup the problem following \cite{gardiner05}: the computational 
domain is two dimensional with $x \in [-1, 1]$, $y \in [-0.5,0.5]$ 
using periodic boundary conditions. Density and pressure are uniform 
with $\rho=1$ and $P = 1$. The particles are laid down in a cubic 
lattice configuration with velocity initialized according to 
${\bf v} = (v_{0}\cos{\theta}, v_{0}\sin{\theta})$ with $\cos{\theta} 
= 2/\sqrt{5}$, $\sin{\theta} = 1/\sqrt{5}$ and $v_{0}=1$ such that by 
$t=1$ the field loop will have been advected around the computational 
domain once. The magnetic field is two dimensional, initialized using 
a vector potential given by
\begin{equation}
A_{z} = \left\{ \begin{array}{ll}
A_{0}(R-r) & r \le R, \\
0 & r > R,
\end{array}\right.
\end{equation}
where $A_{0} = 10^{-3}$, $R=0.3$ and $r = \sqrt{x^{2} + y^{2}}$. The 
ratio of thermal to magnetic pressure is thus given by 
$\beta_{\rm plas} = P/(\frac12 B^{2}) = 2 \times 10^{6}$ (for $r < R$) such 
that the magnetic field is passively advected. 
\cite{gardiner05} show the results of this problem after two crossings of 
the computational domain, by which time the loop has either been 
significantly diffused or has disintegrated into oscillations depending 
on details of their particular choice of scheme. The advantages of a 
Lagrangian scheme are that advection is computed exactly, and using our 
Euler potential formulation for the magnetic field (which in two 
dimensions is equivalent to a vector potential formulation with 
$\alpha = A_{z}$ and $\beta = z$), this is also true for the evolution 
of the magnetic field. The result is that the field loop is advected 
\emph{without change} by our code for as long as one may care to 
compute it. This is demonstrated in Fig.~{\ref{fig:jadvect2D} which 
shows the magnetic field lines at $t=0$ (top) and after 1000 (!) 
crossings of the computational domain (bottom), in which the field 
configuration can be seen to be identical to the top figure. The magnetic 
energy (not shown) is also maintained exactly, whereas \cite{gardiner05} find 
of order a 10\% reduction in magnetic energy after two crossings of 
the domain.\\
In a realistic simulation involving MHD shocks there will be some 
diffusion of the magnetic field introduced by the addition of artificial 
diffusion terms, which are required to 
resolve discontinuities in the magnetic field. However the point is 
that these terms are explicitly added to the SPH calculation and can 
be turned off where they are not necessary whereas the
diffusion present in a grid-based code is intrinsic and always present. \\


{\em 2D: Orszag-Tang test}\\
The evolution of the compressible Orszag-Tang vortex system \cite{orszag79} 
involves the interaction of several shock waves traveling at different 
speeds. Originally studied in the context of incompressible MHD turbulence,
it has later been extended to the compressible case
\cite{dahlburg89,picone91}. It is generally considered a good test 
to validate the robustness of numerical MHD schemes. In the SPH context, this
test has been discussed in detail by \cite{price04c} and \cite{price05}.\\
The problem is two dimensional with periodic boundary conditions on the 
domain $[0,1] \times [0,1]$. The setup consists of an initially 
uniform state perturbed by periodic vortices in the velocity field, which, 
combined with a doubly periodic field geometry, results in a complex 
interaction between the shocks and the magnetic field.\\
The velocity field is given by $\vec{v} = v_{0}[-\sin{(2\pi y)}, 
\sin{(2\pi x)}] $ where $v_{0} = 1$. The magnetic field is given by 
$\vec{B} = B_{0}[-\sin{(2\pi y)}, \sin{(4\pi x)}]$ where $B_{0} = 
1/\sqrt{4\pi}$. Using the Euler potentials this corresponds to 
$\alpha \equiv A_{z} = B_{0}/(2\pi) [ \cos{(2\pi y)} + 
\frac12\cos{(4\pi x)}]$. The flow has an initial average Mach number 
of unity, a ratio of magnetic to thermal pressure of $10/3$ and we 
use a polytropic exponent $\gamma = 5/3$. The initial gas state is therefore $P
= 5/3 B_{0}^{2}= 5/(12\pi)$ and $\rho = \gamma P/v_{0} = 25/(36\pi)$. Note
that the choice of length and time scales differs slightly between various 
implementations in the literature. The setup used above follows that 
of \cite{ryu95} and \cite{londrillo00}.\\
\begin{figure*}
\begin{flushleft}
\includegraphics[height=0.7\textheight,angle=-90]{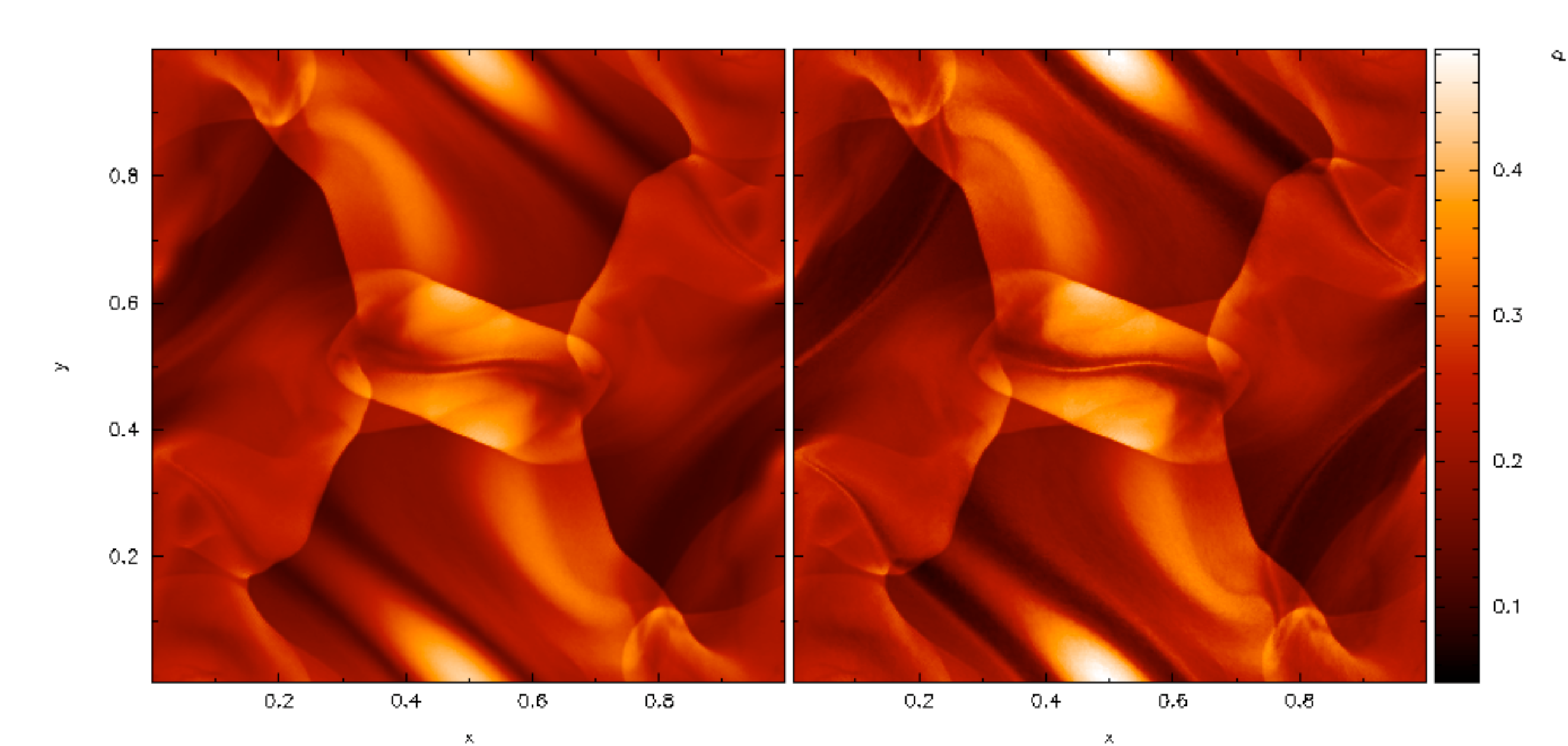}
\caption{Density distribution in the two dimensional Orzsag-Tang 
vortex problem at $t=0.5$. The initial vortices in the velocity 
field combined with a doubly periodic field geometry lead to a complex 
interaction between propagating shocks and the magnetic field. Results 
are shown using $512\times 590$ particles using a SPMHD formalism of
\cite{price05} (left) and using the Euler potentials (right). The 
reduced artificial resistivity required in the Euler potential formalism 
leads to a much improved effective resolution.}
\label{fig:orszagtang}
\end{flushleft}
\end{figure*}
We compute the problem using $512\times 590$ particles initially placed on 
a uniform, close-packed lattice. The density at $t=0.5$ is shown in 
Figure~\ref{fig:orszagtang} using both the SPMHD formalism of \cite{price05}
(left), and the Euler potential approach (right) outlined in 
Sec.\ref{StephanRosswog:sec:Euler}. The Euler potential 
formulation is clearly superior to the standard SPMHD method. This is 
largely a result of the relative requirements 
for artificial resistivity in each case. In the standard SPMHD method the 
application of artificial resistivity is crucial for this problem (that is, 
in the absence of artificial resistivity the density and magnetic field 
distributions are significantly in error). Using the Euler potentials we 
find that the solution can be computed using zero artificial resistivity, 
relying only on the ``implicit smoothing'' present in the computation of 
the magnetic field using SPH operators. This means 
that topological features in the magnetic field are much better preserved, 
which is reflected in the density distribution. For example the filament 
near the center of the figure is well resolved using the Euler potentials 
but completely washed out by the artificial resistivity in the standard 
SPMHD formalism. Also the high density features near the top and bottom 
of the figure (coincident to a reversal in the magnetic field) are much 
better resolved using the Euler potentials.\\

{\em 3D: MHD blast wave}\label{sec:MHD_blast}\\
\begin{figure*}
\begin{center}
\includegraphics[height=0.5\textheight,angle=-90]{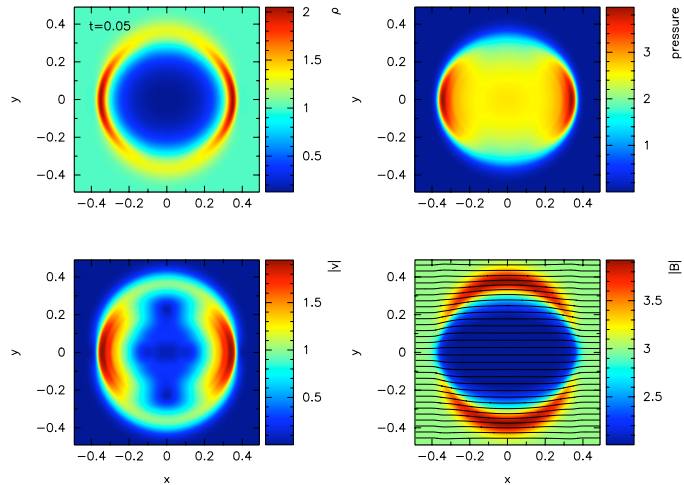}
\caption{Results of the 3D MHD blast wave test at $t=0.05$ at a 
resolution of 1 million ($100^{3}$) particles. Plots show (left 
to right, top to bottom) density, pressure, magnitude of velocity 
and magnetic field strength (with overlaid field lines), plotted in 
a cross-section slice through the $z=0$ plane.}
\label{fig:sedovmhd}
\end{center}
\end{figure*}
 The MHD version of the Sedov test is identical to the hydrodynamic 
test with the addition of a uniform magnetic field in the $x-$direction, 
that is ${\bf B} = [B_{0},0,0]$ with $B_{0} = 3.0$. Initially the 
surrounding material has zero thermal pressure, meaning that the 
plasma $\beta_{\rm plas}$ is zero (ie. magnetic pressure infinitely strong 
compared to thermal pressure). However, this choice of field strength 
gives a mean plasma $\beta_{\rm plas}$ in the post-shock material of $\beta_{\rm plas}\sim 1.3$, 
such that the magnetic pressure plays an equal or dominant role in the 
evolution of the shock.  The results of this problem at $t=0.05$ are 
shown in Fig.~\ref{fig:sedovmhd}, where plots show density, pressure, 
magnitude of velocity and magnetic field strength  in a 
cross section slice taken at $z=0$. In addition the magnetic field 
lines are plotted on the magnetic field strength plot. \\
In this strong-field regime, the magnetic field lines are not 
significantly bent by the propagating blast wave but rather strongly 
constrain the blast wave into an oblate spheroidal shape. The density 
(and likewise pressure) enhancement in the shock is significantly 
reduced in the $y-$direction (left and top right panels) due to the 
additional pressure provided by the magnetic field which is compressed 
in this direction (bottom right panel).

\section{Summary and conclusion}
We have outlined several recent developments in smooth particle hydrodynamics.
The equations of self-gravitating, ideal hydrodynamics were derived 
explicitely from a Lagrangian thereby yielding the correct particle 
index symmetries that ensure that the physical conservation laws are
hard-wired into the discrete set of SPH equations without any arbitrariness.
We have further described the implementation of ideal MHD via so-called Euler
potentials. This approach enforces the crucial 
$\nabla\cdot\vec{B}=0$-constraint 
by construction. 
All dissipative terms required to capture discontinuities were
carefully designed so that they a) have a form suggested in analogy
with Riemann-solvers and b) are only active near discontinuities.
These principles are implemented in our three-dimensional,
Lagrangian magnetohydrodynamics code MAGMA. In a set of standard
test problems used to benchmark numerical (magneto-)hydrodynamics schemes 
we have demonstrated the excellent performance of the code.\\

{\bf Acknowledgement}\\
DJP is supported by a UK Royal Society University Research Fellowship 
though much of this work has been funded by a PPARC/STFC postdoctoral  
fellowship.\\
Some of the results were visualized using SPLASH \cite{price07d},
a publicly available visualisation tool for SPH.

\end{document}